\documentclass[a4paper,fleqn,usenatbib]{mnras}

\usepackage{newtxtext,newtxmath}
\usepackage[T1]{fontenc}
\usepackage{ae,aecompl}

\usepackage{graphicx}	% Including figure files
\usepackage{amsmath}	% Advanced maths commands
\usepackage{amssymb}	% Extra maths symbols
\usepackage{lscape}

\title[ ]{The first \textit{insitu} observation of torsional Alfv\'{e}n waves during the interaction of large-scale magnetic clouds.}

\author[Anil Raghav \& Ankita Kule]{
Anil N. Raghav,$^{1}$\thanks{E-mail: raghavanil1984@gmail.com}
 and Ankita Kule,$^{1}$
\\
$^{1}$University Department of Physics, University of Mumbai, Vidyanagari, Santacruz (E), Mumbai-400098, India\\
}

\date{Accepted XXX. Received YYY; in original form ZZZ}

\pubyear{2017}

\begin{document}
\label{firstpage}
\pagerange{\pageref{firstpage}--\pageref{lastpage}}
\maketitle

\begin{abstract}
The large-scale magnetic cloud such as coronal mass ejections (CMEs) is the fundamental driver of the space weather. The interaction of the multiple-CMEs in interplanetary space affects their dynamic evolution and geo-effectiveness. The complex and merged multiple magnetic clouds appear as the \textit{insitu} signature of the interacting CMEs. The Alfv\'{e}n waves are speculated to be one of the major possible energy exchange/dissipation mechanism during the interaction.  However, no such observational evidence has been found in the literature. The case studies of CME-CME collision events suggest that the magnetic and thermal energy of the CME is converted into the kinetic energy. Moreover, magnetic reconnection process is justified to be responsible for merging of multiple magnetic clouds. Here, we present unambiguous evidence of sunward torsional Alfv\'{e}n waves in the interacting region after the super-elastic collision of multiple CMEs. The Wal\'{e}n relation is used to confirm the presence of Alfv\'{e}n waves in the interacting region of multiple CMEs/magnetic clouds. We conclude that Alfv\'{e}n waves and magnetic reconnection are the possible energy exchange/dissipation mechanisms during large-scale magnetic clouds collisions. The present study has significant implications not only in  CME-magnetosphere interactions but also in the interstellar medium where interactions of large-scale magnetic clouds are possible.
\end{abstract}


%%%%%%%%%%%%%%%%%%%%%%%%%%%%%%%%%%%%%%%%%%%%%%%%%%

%%%%%%%%%%%%%%%%% BODY OF PAPER %%%%%%%%%%%%%%%%%%

\section{Introduction}
The Coronal mass ejections (CMEs) are frequent discharge of huge energy and massive magnetized plasma from the solar corona into the heliosphere. They are of paramount importance in space physics for their key role in extreme space weather and geo-effectiveness \textit{e.g.} \citep{schrijver2010heliophysics,JGRA:JGRA11550,cannon2013extreme,low2001coronal}. 
In last few decades, the understanding of CMEs improved significantly because of space and ground-based observational data with the help of various modeling efforts. The studies are focused on the morphological and kinematic evolution of CMEs in the heliosphere \textit{e.g.} \citep{lindsay1999relationships,st2000properties,chen2011coronal,webb2012coronal,lugaz2017interaction,wang2016twists,zurbuchen2006situ}. By considering the number of CMEs emitted from the Sun during solar maximum and variations in their respective speeds, the interaction between multiple CMEs in the heliosphere is expected to be more frequent.  The collision of multiple CMEs highly affect their dynamic evolution properties and contribute to enhanced geo-effectiveness \textit{e.g.} \citep{wang2005mhd,lugaz2005numerical,lugaz2012deflection,xiong2007magnetohydrodynamic,temmer2012characteristics,shen2011three,shen2012super,wang2002multiple,farrugia2004evolutionary}. To predict space weather effects near the Earth, an accurate  estimation of  arrival time of CMEs at the Earth
is crucial \citep{mishra2014morphological}. Besides this, the study of CME-CME  and CME-solar wind interactions provide unique observational evidences to understand energy dissipation of large-scale magnetic clouds in interstellar medium and authenticate the physical processes predicted theoretically. Therefore, interaction of multiple CMEs needs to be examined in detail.
The various results obtained from studies have justified CME-CME collision as an in-elastic/elastic collision or super-elastic collision \textit{e.g.} \citep{lugaz2012deflection,shen2012super,mishra2017assessing,shen2016turn,lugaz2017interaction}. The magnetohydrodynamics (MHD) numerical
simulations have striven to understand the physical mechanism involved in CME-CME interaction, CME-CME driven shock interactions and their consequences \textit{e.g.} \citep{shen2016turn,niembro2015analytical,jin2016numerical,wu2016numerical}. 

The interaction of multiple CMEs and/or their interaction with any other large-scale magnetic solar wind structures can modify their structural configuration.  The first possibility of CME-CME interaction was reported by analyzing \textit{insitu} observations of CMEs by
the Pioneer 9 spacecraft \citep{intriligator1976august}. \citet{burlaga1987compound} showed that compound streams are formed due to CME-CME interaction using \textit{insitu} observations of the twin Helios spacecraft. \citet{burlaga2002successive} inferred that a set of successive halo CMEs, merged en route from the Sun to the Earth and formed complex ejecta in which the identity of individual CMEs was lost \citep{burlaga2001fast}. Furthermore, CME-CME interactions led to the magnetic reconnection between flux ropes of CMEs and appeared as  multiple magnetic clouds in \textit{insitu} observations \textit{e.g. } \citep{wang2002multiple,wang2003multiple,marivcic2014kinematics,farrugia2004evolutionary}. This mechanism is also known to lead to solar energetic particle (SEP) events \citep{gopalswamy2002interacting}.

It is expected that in the interaction of large-scale magnetic structures, the transfer of momentum and energy takes place in the form of magneto-hydrodynamic (MHD) waves \citep{jacques1977momentum}. Despite this fact, the MHD waves are not commonly observed within the magnetic clouds of various sizes in the solar wind. \citet{gosling2010torsional} reported the first observation of torsional Alfv\'{e}n wave embedded in the magnetic cloud. The Alfv\'{e}n wave fluctuations in the solar wind is a common observable feature \textit{e.g.} \citep{yang2016observational,marubashi2010torsional,zhang2014alfven}. The Alfv\'{e}nic fluctuations are observed in the region where fast and slow solar wind streams interact \textit{e.g.} \citep{tsurutani1995large,lepping1997wind}. Observations also suggest the presence of the Alfv\'{e}n waves during interface of magnetic cloud and solar wind stream   \citep{lepping1997wind,behannon1991structure}. Here, we present the first observation of  Alfv\'{e}n waves embedded in multi-cloud, complex interacting region caused by interaction of multiple CMEs. %The Alfv\'{e}n waves are thought to permeate many astrophysical plasmas. They have been directly observed in the solar wind and  are  expected to exist in the interstellar medium on many length scales and in many environments \citet{goldstein1978instability}.

\section{Methods and Observations}
The multiple-CMEs collision event under study has been studied in the past;  by focusing on (i) their interaction corresponding to different position angles \citep{temmer2014asymmetry} (ii) their geometrical properties and the coefficient of restitution for the head-on collision scenario \citep{mishra2014morphological}. Mari\v{c}i\'{c} et al. (2014) studied heliospheric and \textit{in situ} observations of the same event to understand the corresponding Forbush decrease phenomenon \citep{marivcic2014kinematics,raghav2017forbush,raghav2014quantitative}. It was inferred  that there was a combination of three CMEs instead of two CMEs. Those three CMEs ejected on $13^{th} $ (here onward designated as CME1), $14^{th}$ (CME2) and $15^{th}$ (CME3) February 2011 interacted on their way and appeared as a single complex interplanetary disturbance at 1 AU in the WIND satellite data on 18/20 February 2011 \citep{marivcic2014kinematics,vrvsnak2007transit}.

\begin{figure*}
\begin{center}

\includegraphics[width=1\textwidth]{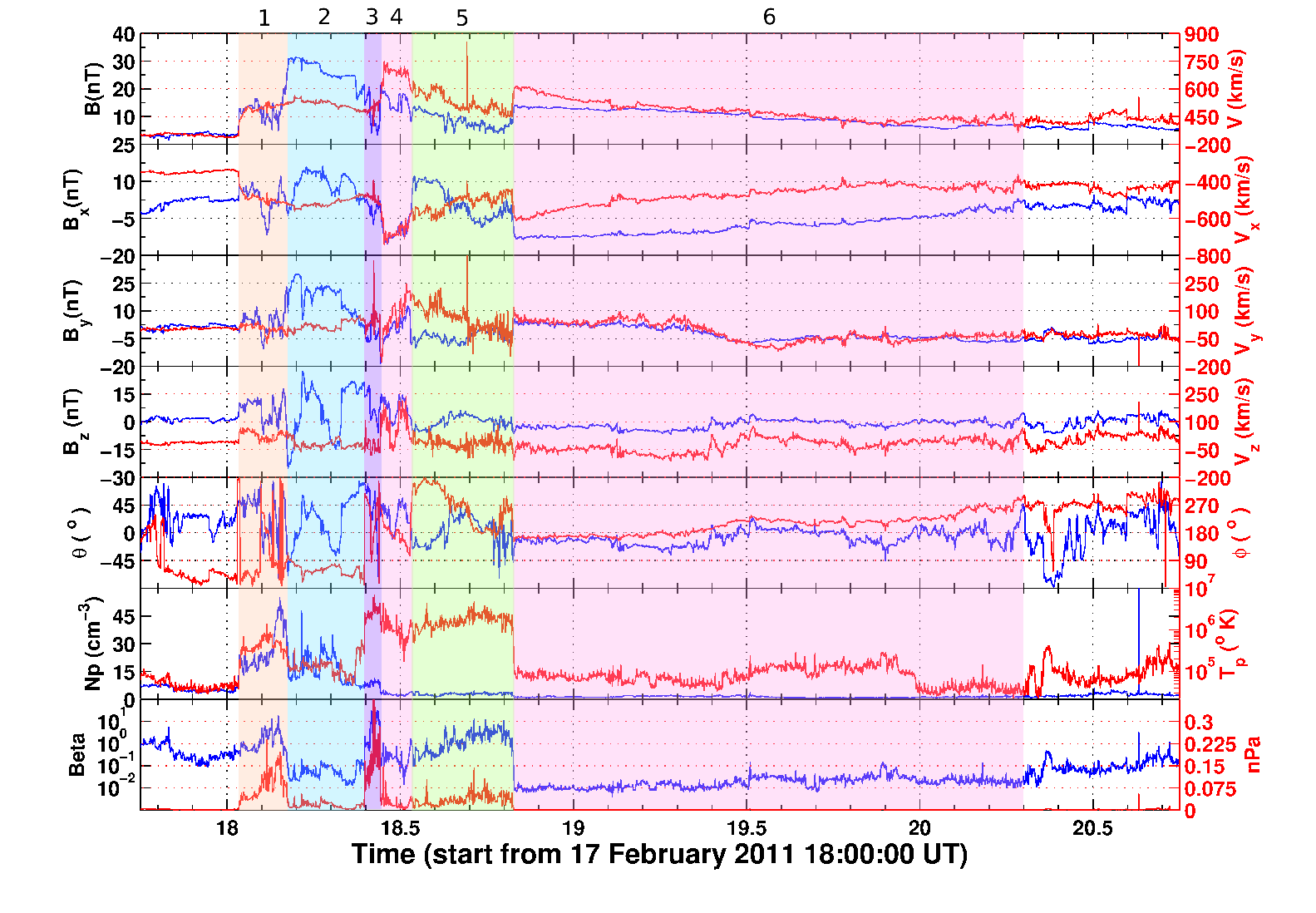}
\caption{Wind observation of complex CME-CME interaction event crossed on 18-20 February 2011 ( time cadence of 92 sec). The top panel shows total interplanetary field strength IMF ($|B|$) and total solar wind $(V)$. The $2^{nd}$, $3^{rd}$ and $4^{th}$ panel from top show IMF components $(B_z, B_y, B_x)$ and solar wind components $(V_z, V_y, V_x)$ respectively. The fifth panel shows IMF orientation $(\Phi, \Theta)$. The sixth panel shows plasma proton density and temperature and bottom panel show plasma beta and plasma thermal pressure. All observations are in GSE coordinate system. The sub-regions of the complex event are presented as a number given at the top and different color shades for better understanding of \textit{in situ} data.
}
\label{fig:1}

\end{center}
\end{figure*}

The \textit{in situ} observations of the highly complex structure is illustrated in Figure \ref{fig:1} consisting of several different regions, which have been marked by numbering on the top with different color shades. The first (reddish-yellow shade) region shows clear sharp discontinuity in all plasma and magnetic field data, which is interpreted as an onset of shock. In general, the presence of the shock should be confirmed with Rankine-Hugoniot relation. The CfA Interplanetary Shock Database available at \url{https://www.cfa.harvard.edu/shocks/wi_data/00530/wi_00530.html} validates the observations. The shock-front is followed by high plasma density, temperature and thermal pressure; large magnetic field fluctuations and enhanced magnetic field strength which is manifested as shock sheath region \citep{richardson2011galactic}. In the region 2 (cyan shade), the elevated magnetic field strength, a decrease in plasma temperature and thermal pressure, very low plasma beta ($\beta$) and gradual decrease in solar wind speed is observed which is ascribed to a magnetic cloud like-structure \citep{burlaga1982magnetic,zurbuchen2006situ}. However, due to the interaction with 2nd CME, the clear signature of the magnetic field rotation is not evident.  

Region 3 (purple shade) shows a sharp drop in the magnetic field strength, increase in plasma parameters such as plasma temperature, plasma beta ($\beta$), thermal pressure and turbulent nature of solar wind speed. This region is sandwiched between two sharp discontinuities. The demonstrated variations in the magnetic field and plasma parameters indicate the presence of the magnetic re-connection-outflow exhaust \citep{gosling2005magnetic,gosling2005direct,xu2011observations} between two CME magnetic clouds \citep{wang2003multiple,lugaz2017interaction}. The detailed justification for this ad-hoc hypothesis is presented in \citet{marivcic2014kinematics} based on similar investigation of \textit{in-situ} data. Region 4 and 5 (pink and green shade) show increase in solar wind speed, lower proton density, gradual decrease in magnetic field strength. The plasma beta is decreasing in regions 4 whereas gradually increasing in region 5. The magnetic field rotation is observed at the beginning of the region 5, but it is not evident during the complete regions of 4 and 5. These observations suggest the highly complex interplanetary region. 

Region 6 (pink shade) shows a sudden increase in magnetic field strength and solar wind speed with steep drop in plasma temperature and proton density, which appears like a shock or a magnetic cloud front boundary after sheath region. Furthermore, $B_z$ shows almost no rotation, $B_y$ shows partial rotation. The $\Theta$ angle is more or less constant whereas $\Phi$ varies by about 45 degrees. This manifested that the region 6 is indeed a magnetic cloud or magnetic cloud-like structure, but one with relatively small rotation. It is listed in the Lepping MC list available at \url{https://wind.gsfc.nasa.gov/mfi/mag_cloud_S1.html} .

The regions 4 and 6 (pink shade) show peculiar and distinct feature in which similar temporal variations are observed in the respective magnetic field and velocity components. The observations indicate the presence of possible magneto-hydrodynamic plasma oscillations \textit{i.e.}  torsional Alfv\'{e}n waves. Typically, two analysis methods are used to confirm the presence of Alfv\'{e}n waves in solar wind/magnetic clouds. In the first method, one can find Hoffman-Teller (HT) frame velocity $(V_{HT})$ using the measured values of B and V. The strong correlation between components of $(V-V_{HT})$ and Alfv\'{e}n velocity confirms the presence of Alfv\'{e}n waves \citep{gosling2010torsional}. However, here we used another obvious characteristic for identifying Alfv\'{e}n waves. The well-correlated changes in magnetic field $B$ and plasma velocity $V$ which is described by the Wal\'{e}n relation \citep{walen1944theory,hudson1971rotational} as 
\begin{equation}
 V_A = \pm A~ \frac{B}{\sqrt{\mu_0 \rho}}
\end{equation}

where $A$ is the anisotropy parameter, $B$ is magnetic field vector and $\rho$ is proton mass density. In the solar wind, the influence of the thermal anisotropy is often not important and can be ignored, thus we usually take $A=1$ \citep{yang2016observational}. The fluctuations $\Delta B$ in $B$ are obtained by subtracting average value of $B$ from each measured values. Therefore, the fluctuations in Alfv\'{e}n velocity is 
\begin{equation}
\Delta V_A = \frac{\Delta B}{\sqrt{\mu_0 \rho}}
\end{equation}

\begin{figure*}
\begin{center}
\includegraphics[width=1 \textwidth]{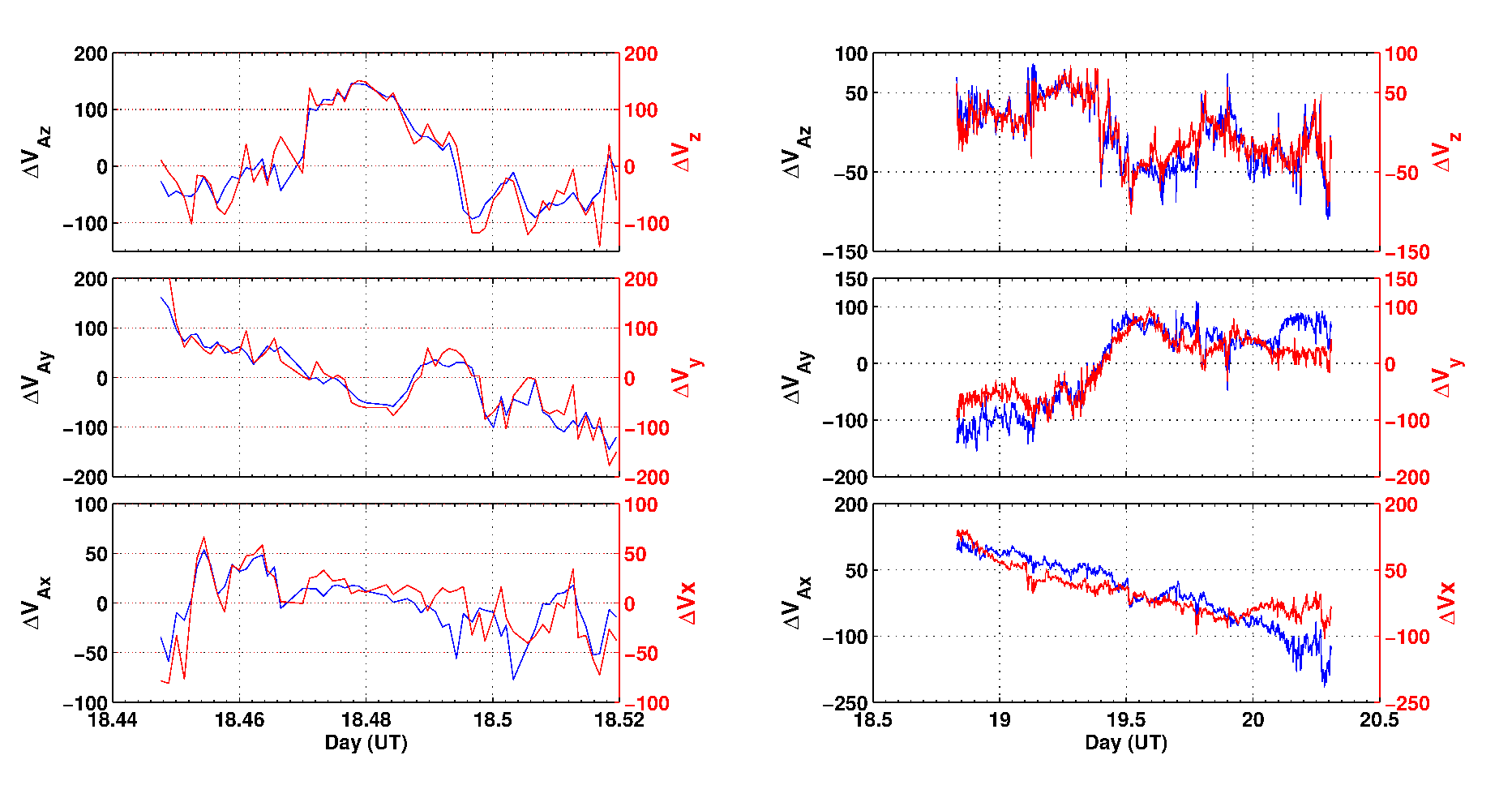}
\caption{Right 3 panels (region 4) and left 3 panels (region 6) illustrate relative fluctuation of Alfv\'{e}n velocity vector $\Delta V_A$ (blue lines) and that of proton flow velocity vector
$\Delta V$ (red lines). ( time cadence of 92 sec) }
\label{fig:2}
\end{center}
\end{figure*}

\begin{figure*}
\begin{center}
\includegraphics[width= \textwidth]{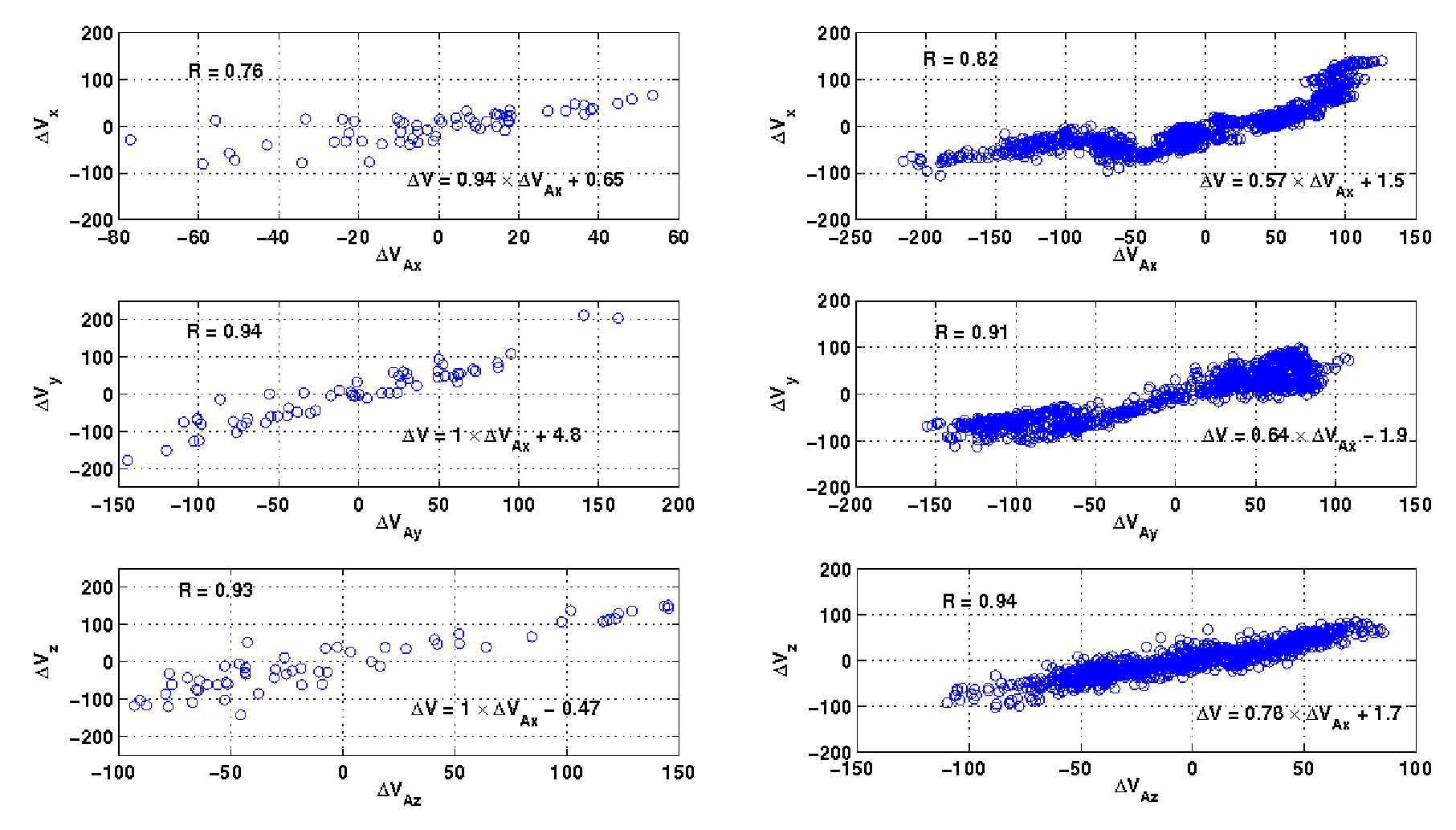}
\caption{The linear relation between $\Delta V_A$ and $\Delta V$ for region 4 (left) and region 6 (right) for the event shown in Figure \ref{fig:1}.   The scattered blue circles are observations from Wind satellite with time cadence of 92 sec. The R is the correlation coefficient. The equation in each panel suggest the straight-line fit relation between respective components of $\Delta V_A$ and $\Delta V$.}
\label{fig:3}
\end{center}
\end{figure*}

Furthermore, the fluctuations of proton flow velocity $\Delta V$ are calculated by subtracting averaged proton flow velocity from measured values. Figure \ref{fig:2} shows the comparison of x, y and z components of $\Delta V_A$ and $\Delta V$, respectively.  Figure \ref{fig:3} shows the linear regression relation between the fluctuations of Alfv\'{e}n velocity vector components and the fluctuations of proton flow velocity vector components for region 4 and region 6 of the event shown in Figure \ref{fig:1}. The linear equation and correlation coefficient are shown in each panel. The correlation coefficients for x, y, and z components are 0.82 (0.76), 0.91 (0.94), and 0.94 (0.93), while the slopes are 0.57 (0.94), 0.64 (1), and 0.78 (1), respectively for region 6 (4); The observed values of correlation coefficients and slopes are consistent with reported studies of Alfv\'{e}n waves \citep{lepping1997wind,yang2016observational}. The correlation and regression analysis of both the regions (shown in Figure \ref{fig:2} ) suggest strong positive correlation between $\Delta V_{Ax}$ \& $\Delta V_x$, $\Delta V_{Ay}$ \& $\Delta V_y$, $\Delta V_{Az}$ \& $\Delta V_z$. This indicates sunward pointing torsional Alfv\'{e}n waves are embedded within both these regions \citep{gosling2010torsional,marubashi2010torsional,zhang2014alfven}. 

\section{Discussion and conclusion}

The heliospheric remote imaging analysis concluded that multiple CMEs interacted in interplanetary space before arriving at 1 AU (position of Wind satellite) \citep{mishra2017assessing,marivcic2014kinematics,temmer2014asymmetry}. The observation manifests that CME1 and CME2 are boosted by CME3 due to its excess speed. The Drag based model suggests region 2 as MC of CME1, combined regions 4 and 5 as MC of CME2 and region 6 as MC of CME3 from Figure \ref{fig:1} \citep{marivcic2014kinematics}. The identification of region 3 as magnetic re-connection-outflow exhaust \citep{gosling2005magnetic,gosling2005direct,xu2011observations,marivcic2014kinematics}  indicates that the magnetic reconnection mechanism is a major interacting mechanism between CME1 and CME2. The middle CME2 MC (regions 4 \& 5) is sandwiched between leading, slower CME1 and fast, following CME3. Therefore, it is highly compressed and overheated. 

The recent work by \citet{mishra2017assessing} considered an oblique collision scenario of two CMEs for the same studied event using the heliospheric remote imaging techniques. The CME1 and CME2 merged together by magnetic reconnection and become visible as first CME after collision and CME3 appeared as second CME in heliospheric imaging data. \citet{mishra2017assessing} estimated the coefficient of restitution (e) as 1.65. The collision leads to an increase in the momentum of first CME by 68\% and a decrease of 43\% in the second CME, in comparison to their values before the collision. Thus, the collision results in an increase of 7.33\% in the total kinetic energy of the CMEs and was interpreted as super-elastic by nature \citep{mishra2017assessing}. Moreover, they also indicate 18 hrs of collision time, after which there is a separation of the two CMEs \citep{mishra2017assessing}. The 18 hrs of collision time indicates the large-scale plasmoid collision is not as simple as solid body collision. Therefore, we hypothesize that the leading edge of CME3 is continuously exerting a force on the trailing edge of CME2. The high distortion with compression and heating of magnetic cloud of CME2 (region 5 of Figure \ref{fig:1}) is evident in \textit{insitu} observation. Thus, magnetic energy of the magnetic cloud of CME2 converts to the kinetic energy of the colliding system, which leads to CME-CME super-elastic collision \citep{shen2012super,mishra2017assessing}. It can also cause a change in the force balance conditions of flux ropes. This induces magneto-hydrodynamic wave in which ions plasma oscillate in response to a restoring force provided by an effective tension on the magnetic field lines. The presence of torsional Alfv\'{e}n waves in region 4 (front of CME2 magnetic cloud) and region 6 ( CME3 magnetic cloud) suggests conclusive evidence of this possible physical mechanism. It implies that the MHD waves are the possible energy and momentum exchange/dissipation mode during large-scale plasmoids interaction. The presence of magnetic reconnection-outflow exhaust at the boundary of region 2 and 4 may have prohibited Alfv\'{e}n waves in region 2. 

The dissipation rate of kinetic energy in clumpy, magnetic, molecular clouds is estimated by \citet{elmegreen1985energy}. Their results indicate that for pressure-equilibrium magnetic field strengths, low-density clouds lose most of their kinetic energy by Alfv\'{e}n wave radiation to the external medium. 
This implies that the presence of sunward torsional Alfv\'{e}n waves in the magnetic cloud of the CME3 causes decreases in kinetic energy. Hence, this lowers down the speed of magnetic cloud of CME3 which further leads to the separation of CMEs. 

The Alfv\'{e}n waves are thought to pervade many astrophysical plasmas. They have been observed in the solar wind \textit{e.g.} \citep{yang2016observational,lepping1997wind,tsurutani1995large} and  are  expected to exist in the interstellar medium on many length scales and in many environments \citep{goldstein1978instability}. It is likely to be one of the major energy exchange/dissipation mode in large-scale magnetic cloud collisions \citep{elmegreen1985energy,jacques1977momentum}. The present study confirms this and demonstrates that torsional Alfv\'{e}n waves is the possible energy dissipation mechanisms during large-scale magnetic cloud collisions. 
 
The energy exchange mechanism in CME and planetary magnetosphere interactions is one of the intriguing problems in space-physics or space-weather studies. Each planetary magnetosphere is considered as one type of plasmoid in the heliosphere, with a particular orientation of magnetic field. 
It is clearly noticeable that the slow-fast solar wind stream interaction, magnetic cloud-solar wind stream interaction and magnetic cloud-cloud interaction give rise to torsional Alfv\'{e}n waves in space plasma. In a condition of opposite magnetic orientation of the plasmoids, the magnetic reconnection is the possible physical mechanism of their interaction \textit{e.g.} geomagnetic storms. The fluctuating $B_z$ fields comprising Alfven waves are expected to cause a type of auroral electro-jet activity called High-Intensity, Long Duration, Continuous AE Activity (HILDCAAs)\citep{tsurutani1987cause}. Their presence may contribute substantially to geo-effectiveness of magnetic cloud-magnetosphere interaction \citep{zhang2014alfven}. Therefore, the interaction of these Alfv\'{e}n waves with the magnetic cloud and planetary magnetosphere should be investigated further.

%\begin{multicols}{2}
%\onecolumn
%\begin{landscape}

%\end{multicols}
%\end{landscape}
%\twocolumn

%\onecolumn

%\twocolumn

  \section*{Acknowledgement}
  We are thankful to WIND Spacecraft data providers (wind.nasa.gov) for  making interplanetary data available. We are also thankful  to Department of Physics (Autonomous), University of  Mumbai, for providing us facilities for fulfillment of this work. AR would also like to thank Wageesh Mishra for valuable discussion on CME-CME interaction.  AR also thanks to Yuming Wang and Solar-TErrestrial Physics (STEP) group, USTC, china \& SCOSTEP visiting Scholar program. AR is also thakful to Gauri Datar.

%\section*{Acknowledgements}

%%%%%%%%%%%%%%%%%%%%%%%%%%%%%%%%%%%%%%%%%%%%%%%%%%



\bibliographystyle{mnras}

%\bibliography{sample} % if your bibtex file is called example.bib


%%%%%%%%%%%%%%%%%%%%%%%%%%%%%%%%%%%%%%%%%%%%%%%%%%

%%%%%%%%%%%%%%%%% APPENDICES %%%%%%%%%%%%%%%%%%%%%

%%%%%%%%%%%%%%%%%%%%%%%%%%%%%%%%%%%%%%%%%%%%%%%%%%

% Don't change these lines
\bsp	% typesetting comment
\label{lastpage}
\end{document}